\documentstyle[epsf,preprint,tighten,aps,%
]{revtex}

\begin{document}

\draft

\preprint{GI-TH-97-10}

\title{What QCD sum rules tell about the rho meson\thanks{Work 
supported by GSI Darmstadt and BMBF.}}

\author{Stefan Leupold$^1$, Wolfram Peters$^1$ and Ulrich Mosel$^{1,2}$ }
\address{$^1$Institut f\"ur Theoretische Physik, Justus-Liebig-Universit\"at
Giessen,\\
D-35392 Giessen, Germany}

\address{$^2$Institute for Nuclear Theory, University of Washington,\\ 
Box 351550, Seattle, WA 98195 }

\date{August 8, 1997}

\maketitle

\begin{abstract}
Using a simple parametrization of Breit-Wigner type for the hadronic side of the 
QCD sum rule for $\rho$ mesons in vacuum as well as in a nuclear medium we explore 
the range of values for the mass 
and the width of the $\rho$ meson which are compatible with the operator 
product expansion. 
\end{abstract}

\section{Introduction} 

Recently the behavior 
of vector mesons in a nuclear surrounding became a lively debated issue. 
This was especially triggered
by the new CERES experiments for S-Au and Pb-Au collisions which show an 
enhancement of the dilepton yield for invariant masses somewhat below the vacuum 
mass of the $\rho$ meson
\cite{ceres1,ceres2,ceres3}. As argued some years ago
by Brown and Rho this enhancement
might be due to the restoration of chiral symmetry \cite{brownrho}. In 
their approach the masses of the vector mesons scale with the quark
condensate, i.e.~drop with rising baryonic density. Thus the $\rho$ peak
in the dilepton spectrum ought to be shifted to lower invariant mass
which indeed might be an explanation for the observed enhancement of
the dilepton yield in that region \cite{cassing,ko,bratkov}. However, also other 
scenarios utilizing the idea of chiral symmetry restoration are possible
which predict a rising $\rho$ mass based on the effect that the $\rho$
becomes degenerate with its chiral partner, the $a1$ meson \cite{pisarski}.

On the other hand, some calculations
based on purely hadronic models gave rise to an alternative picture. 
Due to modifications of the pions in nuclear medium which at least in
vacuum form the most important decay channel of the $\rho$ meson the
spectral function of the $\rho$ might be drastically changed 
\cite{chanfray,herrmann,rapp1}. In addition, collisions of the 
vector meson with nucleons from the Fermi sea also influence the
spectral function of the former \cite{friman,rapp2,klingl97,peters}. 
Both pion modifications and collisions with nucleons at least lead to
a broadening of the $\rho$ peak, if not to completely new structures. 
Hence the enhancement in the dilepton yield might also be explained
within a purely hadronic scenario where a lot of strength is shifted to 
lower invariant mass if the $\rho$ peak becomes much broader basically without
changing its pole position \cite{rapp2}. 

Qualitatively the Brown-Rho scenario was supported
by QCD sum rule analysis which showed a dropping of the $\rho$ mass, {\it if}
the spectral function of the $\rho$ is simply modeled by a $\delta$-function
\cite{hats92,hats95,jinlein}. On the other hand, the 
hadronic models cited above seem to indicate that this approximation might no 
longer be appropriate if the vector meson is placed in a medium e.g.~with the 
saturation density of nuclear matter. A more sophisticated ansatz for the spectral
function of the $\rho$ meson including pion corrections caused by the nuclear
surrounding was explored in \cite{asakawa,asako}. Inserting this
spectral function in the hadronic part of the QCD sum rule basically the same 
dropping of the $\rho$ mass as from the simple pole ansatz was extracted. 
However, in \cite{klingl97}, where also $\rho$-nucleon scattering was taken into 
account, no mass shift but only a large peak broadening was found. It was shown 
there that even then agreement between
the left and the right hand side of the sum rule could be obtained. 
This finding clearly casts some doubt on the quantitative predictive power of the 
QCD sum rule analysis. It seems that only after one has chosen his favorite
hadronic model the sum rule can tell whether this model is in agreement
with the operator product expansion of QCD. 

The purpose of the present paper is to explore in a systematic and as far
as possible model independent way which spectral functions for the $\rho$
meson actually are compatible with QCD sum rules. To do so we choose a
simple Breit-Wigner parametrization for the spectral function treating
the pole position and the on-shell width as toy parameters. Varying these
parameters we calculate the difference of the left and right hand side 
of the sum rule. According to the finding of the two groups cited above
\cite{asako,klingl97} 
we do not expect to find one single set of reasonable parameters but
at least a line or a whole band of parameter pairs. Of course, 
sophisticated hadronic models yield much more complicated spectral functions
than the one we will use here. However, since the spectral
function appears only in an integral in the sum rule 
(cf.~eq.~(\ref{eq:sumrule}) below) we expect that the sum rule is not
sensitive to the details of the modeling of the spectral function, but
only to its gross features. Therefore we believe that our analysis
using a simple Breit-Wigner parametrization is also applicable to more
sophisticated hadronic models. 

In the next section we briefly recapitulate the derivation of the QCD sum rule
for the $\rho$ meson in vacuum as well as in a nuclear medium.  
In section \ref{sec:param} we present our parametrization for the spectral
function. Numerical results are given in section \ref{sec:exam}. Finally
we summarize our results in \ref{sec:concl}. 

\section{Operator product expansion}

We start with the time ordered current-current correlator 
\begin{equation}
  \label{eq:curcur}
\Pi_{\mu\nu}(q) = i \int\!\! d^4\!x \, e^{iqx} \langle T j_\mu(x) j_\nu(0) \rangle 
\end{equation}
where $j_\mu$ is the electromagnetic current of the $\rho$ meson,
\begin{equation}
j_\mu = {1\over 2} \left( \bar u \gamma_\mu u - \bar d \gamma_\mu d \right) \,.  
\end{equation}
In the following we will write down all formulae for the case of finite nuclear
density $\rho_N$. The vacuum case is recovered by simply putting $\rho_N$ to zero. 

At finite baryo-chemical potential Lorentz invariance is broken. From $q^\mu$ and
the baryonic current one can
construct two independent projectors $L_{\mu\nu}(q)$ and $T_{\mu\nu}(q)$ which both 
still satisfy current conservation $q^\mu L_{\mu\nu}(q) = q^\mu T_{\mu\nu}(q) =0$
(cf.~e.g.~\cite{gale}). 
Hence the correlator can be decomposed in the following way:
\begin{equation}
  \Pi_{\mu\nu} = \Pi^T T_{\mu\nu} + \Pi^L L_{\mu\nu} \,.
\end{equation}
The scalar functions $\Pi^T$ and $\Pi^L$ in general depend on $q^2$ and $\vec q^2$ 
where $\vec q$ is defined in the rest frame of the nuclear medium. For 
simplicity we restrict ourselves here to the case of vanishing $\vec q$, 
i.e.~where the $\rho$ meson is at rest relative to the baryonic current. 
In this case $\Pi^T$ and $\Pi^L$ become equal,
\begin{equation}
  \label{eq:defPi}
  \Pi^T(q^2,\vec q^2 =0) = \Pi^L(q^2,\vec q^2 =0) =: \Pi(q^2) \,. 
\end{equation}
Actually, most of the calculations concerning QCD sum rules for vector mesons in 
medium were performed for this special case, 
e.g.~\cite{hats92,hats95,asako,klingl97}. Only recently there appeared growing 
interest in the explicit 
$\vec q$ dependence of the $\rho$ spectral function \cite{friman,rapp2,peters}. 
The traditional approach to QCD sum rules for vector mesons, i.e.~the
pole + continuum ansatz, was recently generalized to finite $\vec q$ by Lee 
\cite{lee97}. A generalization of the approach presented here to finite $\vec q$
is in progress. 

For large space like four-momenta the correlator (\ref{eq:curcur}) and thus 
the scalar function $\Pi$ can be calculated from the 
operator product expansion (OPE) of QCD \cite{shif}. 
Utilizing a subtracted dispersion
relation one can make contact with the time like region by 
(see e.g.~\cite{klingl97})
\begin{equation}
  \label{eq:disp}
\Pi(Q^2) = \Pi(0) + c Q^2 + 
{Q^4 \over \pi} \int\limits^\infty_0\!\! ds \,
{{\rm Im} \Pi(s) \over s^2\,(s + Q^2 -i\epsilon) } 
\end{equation}
where we have introduced $Q^2 =-q^2$ and $c$ is a 
subtraction constant. Now the basic idea
of QCD sum rules is to calculate the l.h.s.~of (\ref{eq:disp}) by the OPE and
the r.h.s.~by a hadronic spectral function. 

Following \cite{hats92,hats93} the OPE for the dimensionless quantity 
\begin{equation}
  \label{eq:rdef}
  R(Q^2) = { \Pi(Q^2) \over Q^2 }  
\end{equation}
including condensates up to dimension 6 is given by
\begin{equation}
  \label{eq:decconR}
R(Q^2) = SC + {m_N \over 4 Q^4}\, A_2 \rho_N 
       - {5 m_N^3 \over 12 Q^6} \,A_4 \rho_N  
\end{equation}
where $SC$ denotes the contribution from the scalar condensates,
\begin{equation}
  \label{eq:scal}
  SC = 
-{1\over 8\pi^2} \left(1+{\alpha_s \over \pi}\right) 
    {\rm ln}\left({Q^2 \over \mu^2}\right) 
+{1\over Q^4} m_q \langle \bar q q\rangle 
+ {1\over 24 Q^4} \left\langle {\alpha_s \over \pi} G^2 \right\rangle 
- {112 \over 81 Q^6} \pi \alpha_s \kappa \,\langle \bar q q\rangle^2  \,.
\end{equation}
The last two contributions
to (\ref{eq:decconR}) arise from the non scalar condensates
$\langle {\cal S} \bar q \gamma_\mu D_\nu q\rangle$ and 
$\langle {\cal S} \bar q \gamma_\mu D_\nu D_\sigma D_\tau q\rangle$
(see \cite{hats92,hats93} for details). 
The medium dependence of the condensates is taken into account in leading order
in the baryon density $\rho_N$ \cite{hats92} (for a review on QCD sum rules in 
nuclear matter cf.~\cite{cohen95}). For the scalar condensates this
yields
\begin{equation}
  \langle \bar q q \rangle =  
\langle \bar q q \rangle_{\rm vac} + {\sigma_N \over 2 m_q} \rho_N
\end{equation}
and
\begin{equation}
  \left\langle {\alpha_s \over \pi} G^2 \right\rangle =
\left\langle {\alpha_s \over \pi} G^2 \right\rangle_{\rm vac} 
- {8\over 9} m_N^{(0)} \rho_N \,.
\end{equation}
We take the following numerical values (cf.~\cite{klingl97}) for \\
-- the strong coupling constant,
$   \alpha_s = 0.36 \,, $ \\
-- the current quark mass,
$ m_q = 7\,{\rm MeV} \,, $ \\
-- the vacuum two quark condensate,
$  \langle \bar q q \rangle_{\rm vac} = (-250\,{\rm MeV})^3 \,,$ \\
-- the vacuum gluon condensate,
$  \left\langle 
{\displaystyle \alpha_s \over \displaystyle \pi} G^2 
\right\rangle_{\rm vac} = 
1.2\cdot 10^{-2}\, {\rm GeV}^4  \,, $  \\
-- the nucleon sigma term,
$   \sigma_N = 45 \,{\rm MeV} \,, $ \\
-- and the nucleon mass in the chiral limit,
$  m_N^{(0)} = 750 \,{\rm MeV} \,. $ \\ 
The factors $A_2$ and $A_4$ are determined from the quark parton distributions. 
We use the values \cite{hats92} $A_2 = 0.9$  and $A_4 = 0.12$. 
Finally $\kappa$ parametrizes the deviation of the four quark condensate from 
the product of two quark condensates, i.e.~from the Hartree approximation. This 
value is still not very well determined. Therefore we will choose different values
for $\kappa$ to check the sensitivity of our results. 
Following \cite{hats92,klingl97,leinw}, respectively, we take 
\begin{equation}
  \kappa = 1,\; 2.36, \; 6  
\label{eq:valkappa}
\end{equation}
and neglect a possible $\rho_N$ dependence of $\kappa$ as is done usually in such 
studies \cite{hats92,hats95,jinlein,asako,klingl97}. This is, however, a point 
that clearly deserves more attention 
since model studies \cite{piekarewicz} show that in the case of finite 
density the RPA excitations out of positive energy states counteract the 
effects of the particle-antiparticle excitations. 

It is important to note that we have not taken into account all condensates up
to dimension 6. Especially, we have neglected higher twist operators since 
they are smaller than the twist two contributions \cite{choi,lee94} and 
their determination from the experiment is not very accurate yet. Their influence
on the OPE for the vector mesons in medium is discussed in \cite{hats95,lee97}. 

The convergence of the OPE can be improved by applying a Borel transformation
to (\ref{eq:disp}). The final result is (see e.g.~\cite{pastar} for details about 
the Borel transformation)  
\begin{eqnarray}
{1\over \pi M^2} \int\limits^\infty_0 \!\! ds \,
{\rm Im} R_{\rm HAD} (s) \, e^{-s/M^2} &=&
{1\over 8\pi^2}\left(1+{\alpha_s\over\pi} \right)
\nonumber \\
&& {}+ {1\over M^4} m_q \langle \bar q q\rangle 
+ {1\over 24 M^4} \left\langle {\alpha_s \over \pi} G^2 \right\rangle 
+ {1\over 4 M^4} m_N A_2 \rho_N 
\nonumber \\
&& {}-{56 \over 81 M^6} \pi\alpha_s \kappa \langle \bar q q\rangle^2 
-{5\over 24 M^6} m_N^3 A_4 \rho_N 
  \label{eq:botr}
\end{eqnarray}
where $M$ denotes the Borel mass and we have introduced the label HAD to stress 
that ${\rm Im} R_{\rm HAD}$ is calculated from a hadronic model. In addition, 
we have absorbed the subtraction constant $\Pi(0)$ appearing in 
(\ref{eq:disp}) into ${\rm Im} R_{\rm HAD}$. It is given by the Landau damping
contribution \cite{hats92} presented in the following section. 

\section{Parametrization of the spectral function } \label{sec:param} 

Following previous works \cite{hats92,jinlein,klingl97} we decompose 
${\rm Im} R_{\rm HAD}$ in three parts, the 
contribution from the $\rho$ meson, the continuum part, and the Landau damping
contribution:
\begin{eqnarray}
  \label{eq:hacola}
  {\rm Im} R_{\rm HAD}(s) = 
\pi F {S(s) \over s} \Theta(s_0-s)  
+{1\over 8\pi}\left(1+{\alpha_s\over\pi} \right) \Theta(s-s_0) 
+\delta(s) {\pi \over 4} \rho_N {1\over \sqrt{k_F^2+m_N^2}} \,.
\end{eqnarray}
The sum rule we will examine in the following is obtained from (\ref{eq:botr})
by inserting the decomposition (\ref{eq:hacola}) and taking the continuum and 
Landau damping contributions to the r.h.s.,
\begin{eqnarray}
{1\over M^2} \int\limits^{s_0}_0 \!\! ds \,
F {S(s) \over s}\, e^{-s/M^2} & = &
{1\over 8\pi^2}\left(1+{\alpha_s\over\pi} \right) 
\left( 1 - e^{-s_0/M^2} \right) - {1 \over 4 M^2} \rho_N {1\over \sqrt{k_F^2+m_N^2}} 
\nonumber \\
&& {}+ {1\over M^4} m_q \langle \bar q q\rangle 
+ {1\over 24 M^4} \left\langle {\alpha_s \over \pi} G^2 \right\rangle 
+ {1\over 4 M^4} m_N A_2 \rho_N 
\nonumber \\
  \label{eq:sumrule}
&& {}-{56 \over 81 M^6} \pi\alpha_s \kappa \langle \bar q q\rangle^2 
-{5\over 24 M^6} m_N^3 A_4 \rho_N   \,.
\end{eqnarray}
If strict vector meson dominance (VMD) is applied then $S$ would be 
the spectral function of the $\rho$ meson (see e.g.~\cite{asako}). 
Since we are examining only a parametrization of $S$ in the following 
we do not have to specify whether we are using VMD or not. 
Anyway we will refer to $S$ as a spectral function.  

The key issue now is the choice for the spectral function $S$. 
Clearly in vacuum the spectral function shows a single peak structure which is,
of course, at the position of the free $\rho$ mass. In
a nuclear medium we have no guideline from experiment how the spectral function
might look like. However, examining the hadronic functions presented in the 
literature (e.g.~\cite{rapp2,klingl97}) we also find
single peak structures, possibly somewhat shifted as compared to the vacuum case.
Thus we will also model our spectral function such that it shows one peak at 
invariant mass $m_\rho$ with width $\gamma$. We will treat these two quantities
as free parameters to figure out which combinations of them are actually compatible 
with OPE. 

We choose the following parametrization 
\begin{equation}
  \label{eq:spec}
S(s) = {1 \over \pi} {\sqrt s \,\Gamma(s) \over (s-m_\rho^2)^2+s\,(\Gamma(s))^2} \,,
\end{equation}
i.e.~we especially neglect a possible $s$ dependence of the mass parameter. 

Since the integral appearing in (\ref{eq:sumrule}) is obviously sensitive to
the behavior of $S(s)$ for small values of $s$ it is important to model the 
threshold behavior of $\Gamma$ in a physically reasonable way. In vacuum the 
spectral function is dominated by the decay of the $\rho$ into two pions. 
Taking into account the phase space as well as the derivative coupling of the
$\rho$-$\pi$-$\pi$ vertex we find the threshold behavior 
\begin{equation}
  \label{eq:thresvac}
  \Gamma_{\rm vac}(s) \sim (s-4m_\pi^2)^{3/2} \,\Theta(s-4m_\pi^2) \,.
\end{equation}
Keeping things as simple as possible we use for the vacuum case 
\begin{equation}
  \label{eq:gammavac}
  \Gamma_{\rm vac}(s) = \gamma \,
\left( {1- {\displaystyle 4 m_\pi^2 \over \displaystyle s} \over 
1- {\displaystyle 4 m_\pi^2 \over \displaystyle m_\rho^2} } \right)^{3/2}
 \,\Theta(s-4m_\pi^2)
\end{equation}
with the constant $\gamma$ being the on-shell width of the vector meson. 
For a more sophisticated parametrization of the spectral function in connection
with QCD sum rules in vacuum we refer to \cite{das87}. 

For finite nuclear density not only the decay into pions but also the scattering
with nucleons influences the spectral function. The latter effect also contributes 
below the two pion threshold. If the Fermi motion of the nucleons is neglected
the threshold for the spectral function of a $\rho$ meson {\it at rest} is now 
given by the mass of one pion since the lightest pair of particles which can
be formed in the $\rho$-nucleon collision is a nucleon and a pion. The threshold
behavior is dominated by the lowest possible partial wave. Without any additional 
constraint from the intermediate state formed in the $\rho$-nucleon collision
we assume it to be an s-wave state. Hence we get
\begin{equation}
  \label{eq:thresmed}
  \Gamma_{\rm med}(s) \sim (s-m_\pi^2)^{1/2} \,\Theta(s-m_\pi^2) 
\end{equation}
and thus 
\begin{equation}
  \label{eq:gammamed}
  \Gamma_{\rm med}(s) = \gamma \,
\left( {1- {\displaystyle m_\pi^2 \over \displaystyle s} \over 
1- {\displaystyle m_\pi^2 \over \displaystyle m_\rho^2} } \right)^{1/2}
 \,\Theta(s-m_\pi^2)  \,.
\end{equation}
If one wants to study the sum rule for very low densities both thresholds 
(one and two pion masses) are relevant. In this case one has to think
about a description how to smoothly switch on the one pion threshold with
increasing density. In the following, however, we will concentrate on the case of
saturation density of nuclear matter. Here, if the width is chosen large enough,
there is already a lot of strength in the spectral function below the two pion
threshold. Hence the two pion threshold does not influence the gross features
of the spectral function for the densities we are interested in. 

We note that there are remarkable exceptions from the single peak 
structure by mentioning two examples: First, if the spectral function of the 
$\rho$ in medium is modeled by
a pion loop where the pions are dressed due to medium effects a second sharp
peak at low invariant mass shows up besides the usual broad $\rho$ peak 
\cite{chanfray,herrmann,asako}. Second, a two pole structure with
two maxima appears if collisions with nucleons from the Fermi sea are 
taken into account for the calculation of the $\rho$ spectral function. This  
additional pole is basically generated by the coupling of the nucleon-$\rho$ 
system to the $N(1520)$ resonance \cite{peters}.\footnote{Actually, the two peak
structure shows up if the spectral function is calculated in lowest order in the
nuclear density. If higher order effects are taken into account the second peak
is washed out to a large extent (see \cite{peters} for details).} 
Such spectral functions
are clearly not covered by our toy spectral function. Therefore strictly speaking,
our parametrization is still model dependent. However we cover a large class 
of models with our ansatz since the sum rule (\ref{eq:sumrule}) is not sensitive
to the details of the spectral function but only to its gross features like the
threshold behavior. 

\section{Examining the sum rule } \label{sec:exam} 

Having set up our formalism we can ask which values for mass $m_\rho$ and
width $\gamma$ are compatible with OPE, i.e.~which set of parameters can 
be inserted in (\ref{eq:spec}) such that the sum rule (\ref{eq:sumrule})
is satisfied. Of course the sum rule is not valid for arbitrary values
of the Borel mass $M$. If $M$ becomes too small the expansion in orders
of $1/M^2$ breaks down. On the other hand for very large values of 
$M$ the contribution from perturbative QCD completely dominates 
the sum rule. In this case
we are no longer sensitive to the parameters we are interested in. 
In \cite{leinw,jinlein} the following recipe was suggested to determine a
reasonable Borel window: The minimal $M$ is determined such that 
the terms of order $o(1/M^6)$ on the r.h.s.~of (\ref{eq:botr}) contribute 
no more than $10\%$ to the total value of the r.h.s. For the maximal $M$ we 
require that the continuum part is not larger than the contribution of
the spectral function to the l.h.s.~of (\ref{eq:botr}), i.e. 
\begin{equation}
  \label{eq:mmax}
\int\limits^\infty_0 \!\! ds \,
{1\over 8\pi}\left(1+{\alpha_s\over\pi} \right) \Theta(s-s_0)  \, e^{-s/M^2} \le 
\int\limits^\infty_0 \!\! ds \,
\pi F {S(s) \over s} \Theta(s_0-s) \, e^{-s/M^2}  \,.
\end{equation}

For given values of $m_\rho$, $\gamma$, $F$, and $s_0$ we calculate the
relative deviation of the l.h.s.~from the r.h.s.~of (\ref{eq:sumrule}) and
average these deviations over the Borel window defined above, schematically
\begin{equation}
\label{eq:diffdef}
  d = \int\limits^{M^2_{\rm max}}_{M^2_{\rm min}} \!\! d(M^2) \, 
\left\vert 1- {\rm l.h.s.}/ {\rm r.h.s.} \right\vert 
/ \Delta M^2   
\end{equation}
with
\begin{equation}
  \label{eq:delm}
  \Delta M^2 = M^2_{\rm max} - M^2_{\rm min}  \,.
\end{equation}
Since we
are mainly interested in the ranges of reasonable values for mass and width
we still have to determine $F$ and $s_0$. We use the following strategy:
For given values of $m_\rho$, $\gamma$, and $s_0$ we fix $F$ by the finite
energy sum rule \cite{hats92},
\begin{equation}
  \label{eq:fesr}
  \int\limits^{s_0}_0 \!\! ds \, {\rm Im} R_{\rm HAD}(s) =
\int\limits^{s_0}_0 \!\! ds \, {1\over 8\pi} \left(1+{\alpha_s\over\pi} \right) \,, 
\end{equation}
hence
\begin{equation}
  \label{eq:detf}
  F = {\displaystyle 
        s_0 {\displaystyle 1\over \displaystyle 8\pi^2} 
\left(1+{\displaystyle \alpha_s\over\displaystyle \pi} \right) 
         - {\displaystyle 1 \over \displaystyle 4} \rho_N 
{\displaystyle 1\over \displaystyle \sqrt{k_F^2+m_N^2}}
      \over \displaystyle \int\limits^{s_0}_0 \!\! ds \, 
{\displaystyle S(s) \over \displaystyle s}     }  \,.
\end{equation}
For given values of $m_\rho$ and $\gamma$, $s_0$ is determined by the requirement 
that $d$ in (\ref{eq:diffdef}) ought to be minimal. 

Now we require two criteria for a pair of parameters $m_\rho$ and
$\gamma$ to hold. First, the difference $d$ as given in (\ref{eq:diffdef})
should be reasonably small. In the following we will explore the region
of parameter pairs which fulfill $d\le 0.2\%$ and $d\le 1\%$, respectively.
Second, the Borel window as defined in (\ref{eq:delm}) 
should be reasonably large. 
In \cite{leinw} a window of $1.4\,$GeV$^2$ was found for the vacuum case
in the traditional sum rule approach (vanishing width). 
As shown in \cite{jinlein} the Borel window shrinks with increasing nuclear
density. At nuclear saturation density the extension of the Borel window was 
found to be $0.9\,$GeV$^2$. For vanishing width these are, of course, optimal 
values and not lower limits. In our calculations we require the window 
$\Delta M^2$ to be larger than $0.9\,$GeV$^2$ for the vacuum case; for nuclear
saturation density we take a value of $0.6\,$GeV$^2$ as a lower limit for 
$\Delta M^2$. 

\begin{figure}[ht]
\centerline{
\epsfxsize=10cm \epsfbox{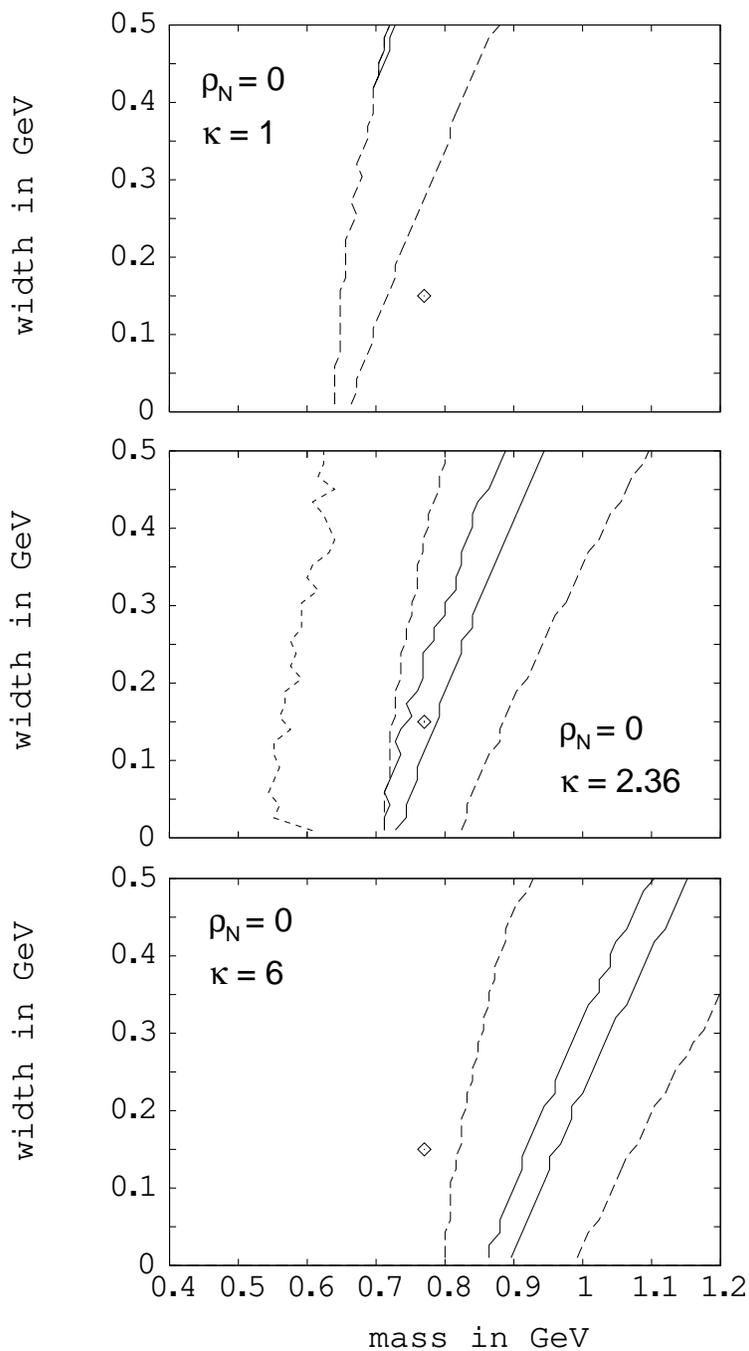}
} 
\caption{\label{fig:vac} The width $\gamma$ over the mass 
$m_\rho$ for vacuum and for different values of $\kappa$. 
The full lines border the region of QCD sum rule allowed 
parameter pairs with $d \le 0.2\%$ and $\Delta M^2 \ge 0.9\,$GeV$^2$, 
the dashed lines border the allowed region 
for $d \le 1\%$ (same $\Delta M^2$). The diamond marks mass and width
of the free $\rho$ meson. The leftmost (dotted) line in the middle picture 
($\kappa=2.36$) gives the
lower boundary for the $d \le 1 \%$ criterion if no constraint on the Borel mass
window is imposed.}   
\end{figure}

\begin{figure}[ht]
\centerline{
\epsfxsize=10cm \epsfbox{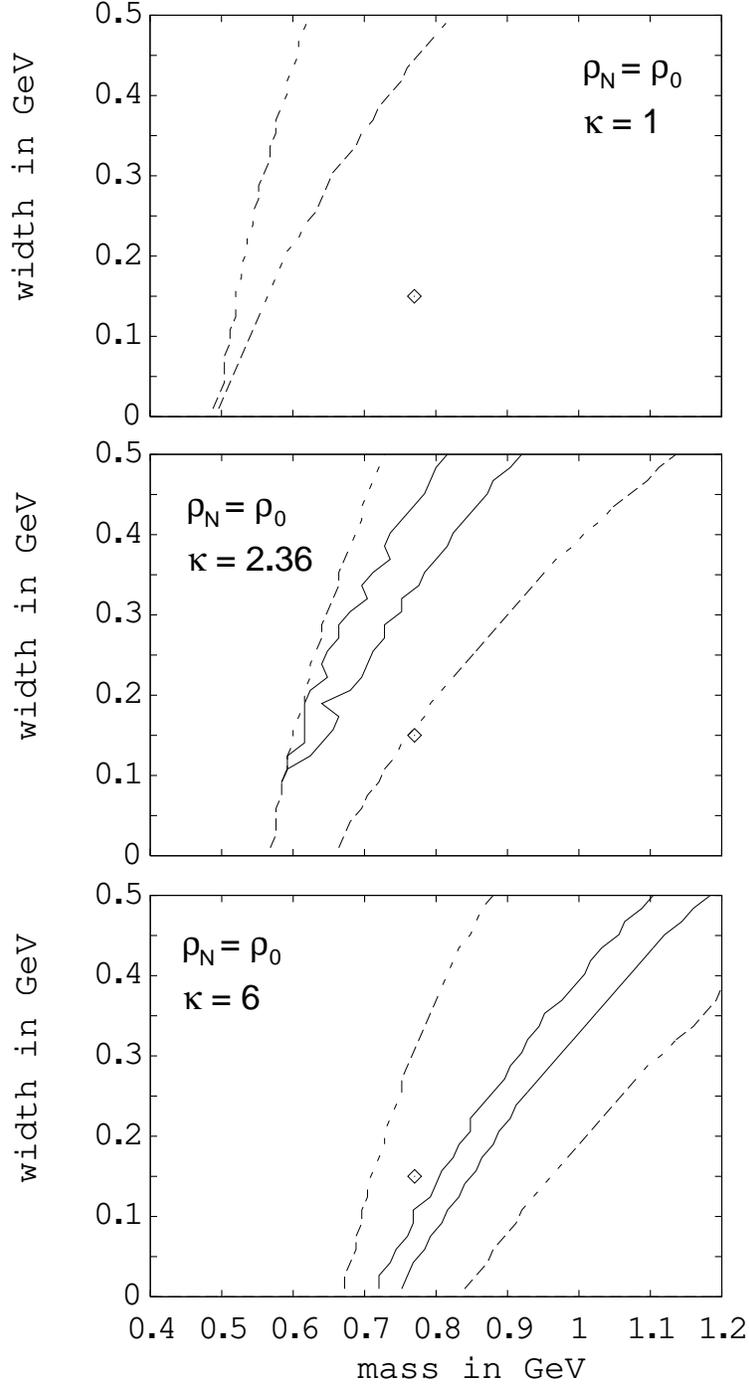}
}
\caption{\label{fig:med} The width $\gamma$ over the mass 
$m_\rho$ for nuclear saturation density $\rho_0$ 
and for different values of $\kappa$. 
The full lines border the region of QCD sum rule allowed 
parameter pairs with $d \le 0.2\%$ and $\Delta M^2 \ge 0.6\,$GeV$^2$, 
the dashed lines border the allowed region 
for $d \le 1\%$ (same $\Delta M^2$). The diamond marks mass and width
of the free $\rho$ meson.} 
\end{figure}

The parameter pairs which fulfill both criteria are plotted in 
figs.~\ref{fig:vac},\ref{fig:med} for vacuum and nuclear saturation density
for the values for $\kappa$ as given in (\ref{eq:valkappa}).  
To get an idea how the two criteria influence the final result we have disentangled
them in fig.~\ref{fig:vac} for the vacuum case and $\kappa=2.36$. 
If no constraint on the width of the Borel window is imposed the left one of 
the dashed lines is replaced by the dotted line. (The other lines stay the same.) 
Obviously
the window criterion gives an important constraint by cutting off low values of
the mass. If the lower 
limit for $\Delta M^2$ is increased the corresponding line (the left dashed one)
would move further to the right. In all the other plots of 
figs.~\ref{fig:vac},\ref{fig:med}
the line which is most to the left is determined by the window criterion. 

For the strict requirement $d \le 0.2\%$ and for $\kappa = 1$ we find only a small 
band of parameters
for the vacuum case and no parameters for the case of nuclear saturation density. 
Note, however, that Hatsuda {\it et al.} 
\cite{hats92,hats93,hats95} who used $\kappa = 1$ took a somewhat larger value 
for the two quark condensate. This would presumably yield a picture similar to 
our case $\kappa = 2.36$, at least for vacuum (fig.~\ref{fig:vac}). 
Actually, only for the choice $\kappa = 2.36$ the experimental values for the 
vacuum case, $m_\rho = 0.77\,$GeV and $\gamma = 0.15\,$GeV, can be found inside 
the parameter range which is in accordance with $d \le 0.2\%$. 
This is however not surprising
since $\kappa = 2.36$ together with the values for the condensates we have used 
here are the result of an optimization to fulfill the sum rule in vacuum reasonably
well \cite{klingl97}. The authors who have used the values $\kappa = 1,\;6$ also use
different values for the other condensates. Clearly, it would be of interest to see 
how a variation
of the other condensate values influences our results. For simplicity we have
restricted ourselves here to a variation of $\kappa$, only. 

The values for $s_0$, $F$, and the Borel window, of course, differ for each pair of
parameters $m_\rho$ and $\gamma$. To get some idea about their magnitudes we
have listed in table \ref{tab:tab} their values for some selected parameter pairs.
One can clearly observe the tendency of the continuum threshold $s_0$ to decrease
with increasing density. This is in agreement with the results from the traditional
sum rule approach \cite{hats92,hats95,jinlein} and was used as an input in the
analysis of \cite{klingl97}. To demonstrate that also the values for $F$ as 
determined from the finite energy sum rule (\ref{eq:fesr}) are in a reasonable
range we take $F = 0.01\,$GeV$^4$ as derived for the vacuum case with 
$\kappa = 2.36$ 
for the physical values $m_\rho = 0.77\,$GeV and $\gamma = 0.15\,$GeV 
(cf.~table \ref{tab:tab}). Assuming vector meson dominance $F$ is indeed given by 
$F \approx m_\rho^4/g_\rho^2 \approx 0.01\,$GeV$^4$ \cite{hats93} where $g_\rho$ 
is the $\rho$-$\pi$-$\pi$ coupling constant. Also F decreases with increasing 
density reflecting the redistribution of $\rho$-strength in the medium.

The results in figs.~\ref{fig:vac},\ref{fig:med} 
are in qualitative agreement with previous work for all values of $\kappa$. 
For vanishing width the mass decreases with increasing density. 
This is the result of the traditional sum rule approach.   
However, with increasing width the window
of ``good'' values for $m_\rho$ is shifted to the right, i.e.~to larger values. 
For finite density this shift becomes even stronger. This confirms the finding
that for a small value of the width the mass drops in medium \cite{asako}, while
for a large value of the width the mass might stay constant \cite{klingl97}. 

With rising width the window
for reasonable masses considerably grows if the weaker criterion $d\le 1\%$ is 
applied. Therefore, the sum rule approach loses
more and more of its predictive power. For e.g.~$\kappa =2.36$ and $d\le 1\%$ we 
find that for small width the mass can be determined reasonably well within an
uncertainty of about $0.1\,$GeV. For large width on the other hand the window
of allowed values for the mass grows to $0.3 - 0.4\,$GeV.  
This in turn means that in any hadronic model that leads to a large width
for the spectral function of the rho meson any agreement of the calculated
mass with the QCD sum rule prediction is not a very strong statement. 
In any case our analysis shows that one cannot extract model 
independent informations about the properties of the $\rho$ meson from the sum 
rules. As already mentioned, 
in all the figures the left one of the dashed lines
is basically determined by the second criterion which requires that the Borel 
window is reasonably large. If this criterion is removed the left one of the 
dashed lines would be placed much more to the left (cf.~fig.~\ref{fig:vac} for
$\kappa = 2.36$). 
Then the uncertainty in the determination of the mass for 
vanishing width would also be much larger than $0.1\,$GeV. 

From the figures for different values of $\kappa$ we find that the masses rise
with growing $\kappa$. At first sight, 
one would tend to exclude the cases $\kappa = 1,\; 6$
for vacuum since the experimental values for mass and width are not inside the
parameter range supported by the sum rule. This, however, does not necessarily
mean that these values for $\kappa$ ought to be discarded for finite density, too. 
Since $\kappa$ might explicitly depend on the density the values we have used are
not unrealistic a priori. 
Fig.~\ref{fig:med} shows that even for vanishing
width a $\kappa$ which grows with the density can compensate for the negative
mass shift otherwise caused by the density and restore the vacuum value of the 
mass. We conclude that the sum rule analysis is still plagued by a large 
uncertainty in the determination of the condensate values, especially concerning
the four quark condensate.

\section{Summary } \label{sec:concl} 

We have presented a systematic study of the QCD sum rule for the $\rho$ meson
to shed some light on the question which
hadronic spectral functions for the $\rho$ meson are compatible with the 
operator product expansion of QCD. Using a simple Breit-Wigner ansatz with 
the correct threshold behavior incorporated we explored the range of parameters
for the $\rho$ mass and on-shell width which satisfy the QCD sum rule. 
From the qualitative point of view the important message is that there is not
only one pair of reasonable parameters. Quantitatively we find that for small
(large) values of the mass also the width has to be small (large). Thus the lowest 
mass is given by
the traditional calculations using a $\delta$-function for the spectral function
\cite{hats92,hats95,leinw,jinlein}. However, especially for large values of the 
width a large range of values for the mass would also be allowed. 
We conclude that specific properties of the
$\rho$ meson in a nuclear medium cannot be determined from QCD sum rules without
specifying a hadronic model. Instead the sum rules only provide a (wide) constraint 
for reasonable hadronic models. Even then, the 
factorization of the four-quark condensate, i.e.~the dependence of any 
result on the parameter $\kappa$ and its possible density dependence, 
introduces an additional uncertainty. 

Nevertheless the sum rules are useful, first, to 
provide a constraint for reasonable hadronic models or, second, to incorporate
additional medium dependences of hadronic parameters (see e.g.~\cite{asako}).  
We stress again, however, that the sum rule loses its predictive power with 
increasing width since the window for values of the mass supported by the 
sum rule drastically grows. For any further study concerning the comparison
of a hadronic model with the OPE we suggest to quantify any statement about
the agreement with the QCD sum rule. We advocate to use the criterion for a 
reasonably large Borel
window as presented by Leinweber {\it et al.} \cite{leinw,jinlein} together with
the deviation $d$ presented here as a tool to check the compliance of any hadronic 
model with the OPE. 

The work presented here is restricted to $\rho$ mesons which are at rest in the
nuclear medium. A generalization to finite momentum is in progress. 

\acknowledgments
The authors thank Carsten Greiner for stimulating discussions. 
One of the authors (U.M.) thanks the Institute for Nuclear Theory at the 
University of Washington for its hospitality and the U.S. Department of 
Energy for partial support during the completion of this work.

\begin{table}[ht]
\begin{tabular}{l|l|l|l|l|l|l|l} 
 & $m_\rho$ [GeV] & $\gamma$ [GeV] & $s_0$ [GeV$^2$] & $F$ [GeV$^4$] & 
$M^2_{\rm min}$ [GeV$^2$] & $M^2_{\rm max}$ [GeV$^2$] & d \\
\hline 
\multicolumn{8}{l}{vacuum, $\kappa = 2.36$} \\ 
\hline 
opt.  & 0.71 & 0.01 & 1.1  & 0.008 & 0.69 & 1.6 & 0.0011 \\ 
\hline
exp.  & 0.77 & 0.15 & 1.2  &  0.01 & 0.69 & 1.8 & 0.0015 \\ 
\hline 
\multicolumn{8}{l}{vacuum, $\kappa = 1$} \\ 
\hline 
opt.  & 0.71 & 0.48 & 0.96 & 0.01  & 0.51 & 1.4 & 0.0013 \\ 
\hline
opt.0 & 0.64 & 0.01 & 0.95 & 0.06  & 0.51 & 1.4 & 0.0070 \\ 
\hline 
exp.  & 0.77 & 0.15 & 1.3  & 0.01  & 0.51 & 1.9 & 0.014 \\
\hline 
\multicolumn{8}{l}{vacuum, $\kappa = 6$} \\ 
\hline 
opt.  & 0.90 & 0.08 & 1.7  & 0.02  & 0.95 & 2.5 & 0.0013 \\ 
\hline 
opt.0 & 0.88 & 0.01 & 1.6  & 0.02  & 0.95 & 2.4 & 0.0013 \\ 
\hline
exp.  & 0.77 & 0.15 & 1.1  & 0.01  & 0.95 & 1.5 & 0.0097 \\ 
\hline 
\multicolumn{8}{l}{nuclear saturation density, $\kappa = 2.36$} \\ 
\hline 
opt.  & 0.66 & 0.22 & 0.88 & 0.006 & 0.54 & 1.3 & 0.00085 \\ 
\hline
opt.0 & 0.57 & 0.01 & 0.78 & 0.003 & 0.54 & 1.1 & 0.0019 \\ 
\hline 
\multicolumn{8}{l}{nuclear saturation density, $\kappa = 1$} \\ 
\hline 
opt.  & 0.62 & 0.48 & 0.68 & 0.005 & 0.41 & 1.0 & 0.0035 \\ 
\hline
opt.0 & 0.49 & 0.01 & 0.67 & 0.002 & 0.41 & 1.0 & 0.0092 \\ 
\hline 
\multicolumn{8}{l}{nuclear saturation density, $\kappa = 6$} \\ 
\hline 
opt.  & 0.74 & 0.01 & 1.2  & 0.009 & 0.74 & 1.7 & 0.0011 \\
\end{tabular}
\caption{\label{tab:tab} Values for $s_0$, $F$, $M^2_{\rm min}$, $M^2_{\rm max}$, 
and $d$ for some selected parameter pairs $m_\rho$ and $\gamma$. 
{\it opt.} denotes the optimal value (minimal $d$) of the whole
parameter range. {\it opt.0} denotes the optimal value for the smallest 
width we have studied. {\it exp.} denotes the experimental values 
$m_\rho = 0.77\,$GeV and $\gamma = 0.15\,$GeV.} 
\end{table} 

\end{document}